\documentclass[11pt,letterpaper]{JHEP3}
\usepackage{graphicx}
\usepackage{epsfig}
\usepackage{amsfonts,amstext,amsmath}
\usepackage{amssymb,latexsym}

\title{Neutrinos in the Simplest Little Higgs Model}
\author{Jaeyong~Lee\\
Department of Physics, Box 1560,
University of Washington, Seattle, WA 98195-1560 \\
E-mail: \email{jaeyong@u.washington.edu}}
\preprint{UW/PT-05-09}

\abstract{The simplest little Higgs model based on a $SU(3)$ global symmetry
contains a $SU(3)_{weak}$ triplet and a singlet per a generation in the lepton sector.
A neutral component of the triplet and the singlet turn into a neutral vector-like
$SU(2)_L$ singlet after electroweak symmetry breaking while the other neutral component
of the triplet is the SM neutrino. At tree level, 
Yukawa couplings of the lepton sector not only allow the neutral vector-like lepton to couple to the SM neutrino, but also give them a Dirac mass.
Majorana mass terms for  the SM neutrinos and their partners arise at one loops,
leading to neutrino flavor mixing in addition to neutrino-heavy neutral lepton mixing.}

\begin{document}

\section{Introduction}

The little Higgs models, as an alternative solution of the ``little hierarchy problem'' of
the Standard Model (SM), revive the idea of the Higgs doublet being a pseudo
Goldstone boson of some global symmetry which is spontaneously broken at a TeV scale.
In the little Higgs models the SM Higgs is a part of the scalar multiplet(s) of some global symmetry
and the SM Higgs mass parameter depends logarithmically on the UV cutoff scale
of the global symmetry.
Among the various little Higgs models, there is the simplest little Higgs model (SLHM)
\cite{Schmaltz:2004de,Schmaltz:2005ky} based on a simple $SU(3)$ global symmetry.
The study of precision data in the gauge and quark sector set
constraints on the parameters of the simplest little Higgs model
\cite{Marandella:2005wd,Casas:2005ev}.

The SLHM contains two distinct features compared with the littlest Higgs model (LHM)
\cite{Arkani-Hamed:2002qy}  which is most intensively examined among the little Higgs models.
The first feature is that the SLHM has no counter partner(s) to the SM Higgs scalar.
The existence of the SM Higgs counter partner in the little Higgs models is required to cancel
dangerous quadratically divergent contribution to the 
SM Higgs mass squared paramters coming from the SM Higgs loops.
In the LHM, the complex $SU(2)_L$ triplet scalar play a role of
the counter partner to the SM Higgs doublet.
However, in the SLHM the absence of the quadratically divergent contribution
to the Higgs mass parameter from the Higgs loops is excused by the fact
that the SM Higgs is contained in a pair of Higgs multiplets
rather than in a Higgs multiplet.

The second feature, barely noticed before, is that the SM neutrino in a generation
accompanies a neutral vector-like lepton
just as the top quark teams up with a heavy vector-like quark.
The heavy top partners lead to the large Yukawa couplings for the top quark.
Unlike the heavy top quark, the SM left-handed neutrinos do not have right-handed pairs.
Furthermore, they have extremely small masses, which has been
convinced by neutrino flavor oscillation experiments
as well as cosmological and astrophysical observations.
The standard formalism for neutrino oscillations is based on oscillations
among the three left-handed neutrinos $(\nu_e,\nu_\mu,\nu_\tau)_L$,
which is refered to as first-class oscillations.
In the SLHM, the coupling of the SM neutrino to its heavy partner further
give rise to transitions between the SM neutrinos and their heavy partners,
which is dubbed as second-class oscillations \cite{Bilenky:1976yj}.

The object of this letter is to make a detailed analysis of the second-class
oscillations as well as the first-class oscillations within the framework of
the SLHM. To do so, we first inspect the origin of the SM neutrino masses
which in general arise from Dirac or Majorana mass terms
in various extensions of the SM.
In the SLHM, though right-handed neutral leptons are present in the
neutral lepton sector they are not Dirac-paired with the SM (left-handed) neutrinos.
Rather they are Dirac-paired with the partners of the SM neutrinos.
Therefore it is naively expected that the SM neutrino mass in the SLHM arises
only from Majorana mass terms. We show that radiative corrections give rise to
the Majorana mass terms, leading to
not only the SM neutrino masses but also the SM neutrino mixing angles.
We further search for other consequences of the scenario at weak energy scale.

The rest of the letter is organized as follows: In section 2 we briefly review
the Higgs sector of the SLHM. In section 3
we discuss the Yukawa couplings of the lepton sector in one generation case.
In section 4 we study the mechanism of the SM neutrino masses in one generation case.
In section 5 we extend the mechanism of the SM neutrino masses to three
generation case and investigate both the first-class and second-class oscillations.
In section 6 we give a summary and provide an outlook for the neutrino physics within the SLHM.
    
\section{Higgs sector}

The simplest little Higgs model based on the $SU(3)$ simple global symmetry
has the initial electroweak gauge structure $SU(3)_{weak}\times U(1)_X$ which is 
broken to the SM $SU(2)_L\times U(1)_Y$.
The symmetry breaking is triggered by the VEV's
of a pair of triplets  $\Phi_{1,2}$, which transform as  $(3,-\frac{1}{3})$
under the $SU(3)_{weak}\times U(1)_X$.
They are parameterized non-linearly  as follows:
\begin{eqnarray}\label{eq:two phis}
\Phi_1=\exp\left\{i\Theta\frac{f_2}{ff_1}\right\}\left(\begin{array}{c}
0 \\ 0 \\ f_1\end{array}\right)&\equiv&
\exp\{i\Theta \cot\beta/f\}\left(\begin{array}{c}
0 \\ 0 \\ f\sin\beta\end{array}\right), \\ \nonumber
\qquad
\Phi_2=\exp\left\{-i\Theta\frac{f_1}{ff_2}\right\}\left(\begin{array}{c}
0 \\ 0 \\ f_2\end{array}\right)&\equiv&
\exp\{-i\Theta\tan\beta/f\}\left(\begin{array}{c}
0 \\ 0 \\ f\cos\beta\end{array}\right)
\end{eqnarray}
where $f_1\equiv f\sin\beta$, $f_2\equiv f\cos\beta$,
and the $\Theta$ NGB matrix is
\begin{equation}\label{eq:vev angle}
\Theta=\frac{\eta}{\sqrt{2}}+\left(\begin{array}
{cc} \begin{array} {cc}0 & 0 \\ 0 & 0 \end{array}  \,\,\,h 
\\ \!\!\!\!\!h^\dagger & \!\!\!\!\!\!0 \end{array}\right). 
\end{equation}    
Here the field $h$ is an $SU(2)_L$ doublet which is identified with
the SM Higgs doublet, $\eta$ is a real $SU(2)_L$ singlet scalar with no vev.
The mass of $\eta$ is assumed to be of the electroweak scale and the complete
analysis of the physics associated with $\eta$ is given in Ref.~\cite{Kilian:2004pp}.

There are yet seven degrees of freedom in $\Phi_1$ and $\Phi_2$
but we have omitted them in Eq.~(\ref{eq:two phis}) by intention. The reason is as follows:
five degrees of freedom are eaten by the gauge fields during the symmetry breaking
of $SU(3)_{weak}\times U(1)_X\to SU(2)_L\times U(1)_Y$ acting as the
longitudinal components of the broken gauge fields
and the rest, two degrees of freedom are the massive excitation modes along the two vev's
directions with their masses being of order $f$.
Thus we can take into account only five pseudo Goldstone bosons, $h$ and $\eta$, below
the scale $f$.

\section{One generation case}

We now take into account the leptonic sector in the SLHM.
For simplicity, we consider only the first generation in this section.
The SM  first leptonic generation is embedded in 
$SU(3)_{weak}\times U(1)_X$ representations as follows:
\begin{eqnarray}
    \Psi_L=(3,-\frac{1}{3})&&\qquad\qquad e_R =(1,1)\\ \nonumber
    &&\qquad\qquad n_R=(1,0)
\end{eqnarray}
where the triplet $\Psi_L=(iL,n_L)^T$ embraces the left-handed
$SU(2)_L$ doublet, $L=(\nu_{eL},e_L)^T$ while the
right-handed $SU(3)_{weak}$ singlets $e_R$ and $n_R$ are the Dirac
partners of $e_L$ and $n_L$ in the triplet, respectively
\footnote{We shall assert that  $n^c_L\neq n_R$
after we introduce the SM neutrino mass.}.
Note that the SM neutrino in the triplet has no Dirac partner
\footnote{There are two different embeddings of the quarks and leptons
in Ref.~\cite{Schmaltz:2004de}. Though the quark 
charges under the 3-3-1 gauge group are different
between the two embeddings, the lepton charges are the same in both.}.

The existence of the two triplet scalars allows us to construct the 
Yukawa couplings, yielding mass terms for the fermions in the model.
In particular, the mass terms for the SM charged lepton $e$ and neutral lepton
$n$ arise from the interactions of the form:
\begin{equation}\label{eq:ecpsi}
-\mathcal L_{\rm yuk}=
\lambda^n \overline {n_R} \Phi^\dagger_1 \Psi_L + 
\frac{\lambda^e}{f} \overline{e_R} \Phi^i_1\Phi^j_2 \Psi_L^k \epsilon_{ijk} 
+h.c.,
\end{equation}
where $\lambda^e$ and $\lambda^n$ are set to be real, which is allowed by
redefining phases of $n_R$ and $e_R$, respectively.  
As in the quark sector,
one may also take into account another term like $\overline{n_R} \Phi_2^\dagger \Psi_L$ 
which is obtained by replacing $\Phi_1^\dagger$ by $\Phi_2^\dagger$
in the first term of $(\ref{eq:ecpsi})$. However, we need not include it because
it does not change the physics associated with neutrinos which we shall discuss.
To make the physics simple, we ignore the interaction from now on.

The second component in the lepton triplet, $e_L$, and the charged singlet,
$e_R$, have a Dirac mass after the SM Higgs (the first component of $\Phi_1$)
gets a vev,
$\langle h^T\rangle =(\frac{v}{\sqrt{2}},0)$:
\begin{equation}
\frac{\lambda^e}{f} \overline{e_R} \Phi^i_1\Phi^j_2 \Psi_L^k \epsilon_{ijk} 
+h.c. \rightarrow -\mathcal L_{\rm Dirac}=
\lambda^e\frac{v}{\sqrt{2}}\overline{e_R}e_L+h.c.,
\end{equation}
yielding the electron mass, $m_e = \lambda^e v/\sqrt{2}$, as in the SM.
On the other hand,  a linear combination of  $\nu_{eL}$ and $n_L$
have Dirac masses via the neutral singlet, $n_R$, as the first
and third components of $\Phi_1$ acquire vevs, $\langle\Phi_1\rangle
=(if_2/f\langle h\rangle,f_1)^T$, respectively:
\begin{eqnarray}\label{eq:nnma}
\lambda^n \overline{n_R} \Phi^\dagger_1 \Psi_L +h.c.
&\rightarrow& \lambda^n f_1\overline{n_R} n_L
+\lambda^n\frac{f_2}{f}\frac{v}{\sqrt{2}} \overline{n_R}\nu_{eL}
+h.c. \nonumber \\
&=& \lambda^n f\overline{n_R}\left(\sin\beta n_L
+\frac{v}{f\sqrt{2}}\cos\beta\nu_{eL}\right)+h.c.
\end{eqnarray}
The relation of mass eigenstate $(\hat\nu_{eL},\hat n_L)$ to
weak eigenstate $(\nu_{eL},n_L)$ is given by
\begin{equation}\label{eq:second osc}
\hat n_L =\cos\theta n_L+ \sin\theta \nu_{eL},\qquad 
\hat \nu_{eL}=-\sin\theta n_L + \cos\theta\nu_{eL}
\end{equation}
with the mixing angle
\begin{equation}\label{eq:mix angle}
\sin\theta\equiv\frac{\frac{v}{\sqrt{2}f}\cos\beta}
{\sqrt{\sin^2\beta+\frac{v^2}{2f^2}\cos^2\beta}},\qquad
\cos\theta\equiv\frac{\sin\beta}{\sqrt{\sin^2\beta+\frac{v^2}{2f^2}\cos^2\beta}}.
\end{equation}
Note that the mixing angle is independent of the Yukawa coupling $\lambda^n$,
and vanishes in the limits of either $v/f\to 0$ or $\beta\to \pi/2$
\footnote{In case that the Yukawa interaction for $n$ is given by
$\lambda^n n^c\Phi_2^\dagger\Psi_L$ instead of $\lambda^n n^c\Phi_1^\dagger\Psi_L$,
the mass $m_n$ is obtained by exchanging
$\sin\beta\leftrightarrow\cos\beta$ in Eq.~(\ref{new_heavy_ma}).}.

The neutral fields $\hat n_L$ and $n_R$ consist of the neutral heavy lepton:
\begin{equation}
-\mathcal L_{\rm Dirac}= m_n\overline{n_R}\hat n_L+h.c.
\end{equation}
with a Dirac mass 
\begin{equation}\label{new_heavy_ma}
m_n=\lambda^n f\sqrt{\sin^2\beta+\frac{v^2}{2f^2}\cos^2\beta},
\end{equation}
which is determined almost by the coupling
$\lambda^n$ because the global symmetry breaking parameter $f$
is assumed to be a TeV range in little Higgs models.
Experimentally, the neutral heavy lepton have a larger mass than the weak gauge bosons,
satisfying the neutral heavy lepton mass limits \cite{Eidelman:2004wy}.
Assuming that the vev parameter is $f\lesssim 10$ TeV we can set a lower bound
on the coupling, $\lambda^n\gtrsim 10^{-2}$.

From Eq.~(\ref{eq:second osc}) we see that second-class oscillations
between $\nu_{eL}$ and $n_L$ can be present so
that the chance of finding the signature of an neutral heavy lepton is proportional
to the mixing angle, $\sin\theta$. But due to a large mass of the neutral heavy lepton
second-class oscillations are not present in neutrino experiments. 
Instead, a way to probe $\sin\theta$,
is to measure decays of the neutral heavy lepton
via charged currents like $n\to W\ell_L$ or neutral currents like $n\to Z\nu_L$.

As for the SM neutrino the Yukawa interactions in Eq.~(\ref{eq:ecpsi})
leaves the SM neutrino to be massless. To explain nonzero mass for the SM neutrino
favored in the neutrino oscillation experiments, we explore not only the origin of the neutrino mass
but also a UV completion of the simplest little Higgs model.
But one should keep in mind that the UV cutoff of little Higgs models is about $10\sim100$ TeV,
which implies that the symmetry breaking scale associated with the SM neutrino masses is so low
that the seesaw mechanism is of no use.
Therefore, we reckon that the SM neutrino masses will stem from another mechanism
which yields a sufficiently small mass for the neutrino.

\section{Mechanism of Majorana masses}
\label{sec:one}
We now turn to our attention to the origin of the SM neutrino masses
in the context of the simplest little Higgs model.
One may think of a Dirac mass term as a source of the SM
neutrino mass due to the presence of the right-handed neutral singlet field, $n_R$.
But we have already shown that
the right-handed singlet does not contributes to a Dirac mass for the SM neutrino.
Rather, it becomes the Dirac pair with the heavy neutral partner of the SM neutrino.
Furthermore, the SM neutrinos in other little Higgs models seems to acquire
a Majorana mass rather than a Dirac mass \cite{Kilian:2003xt,Lee:2005kd,Han:2005nk}.
Accordingly tiny SM neutrino masses in the simplest
little Higgs model are expected to be Majorana,
which was first considered in Ref.~\cite{delAguila:2005yi}.
In what follows we shall further scrutinize the mechanism of
the SM neutrino mass in Ref.~\cite{delAguila:2005yi}.
But for simplicity we consider one generation case in this section
and shall extend to three generation case in Section \ref{sec:3g}.

\FIGURE[l]{\epsfig{file=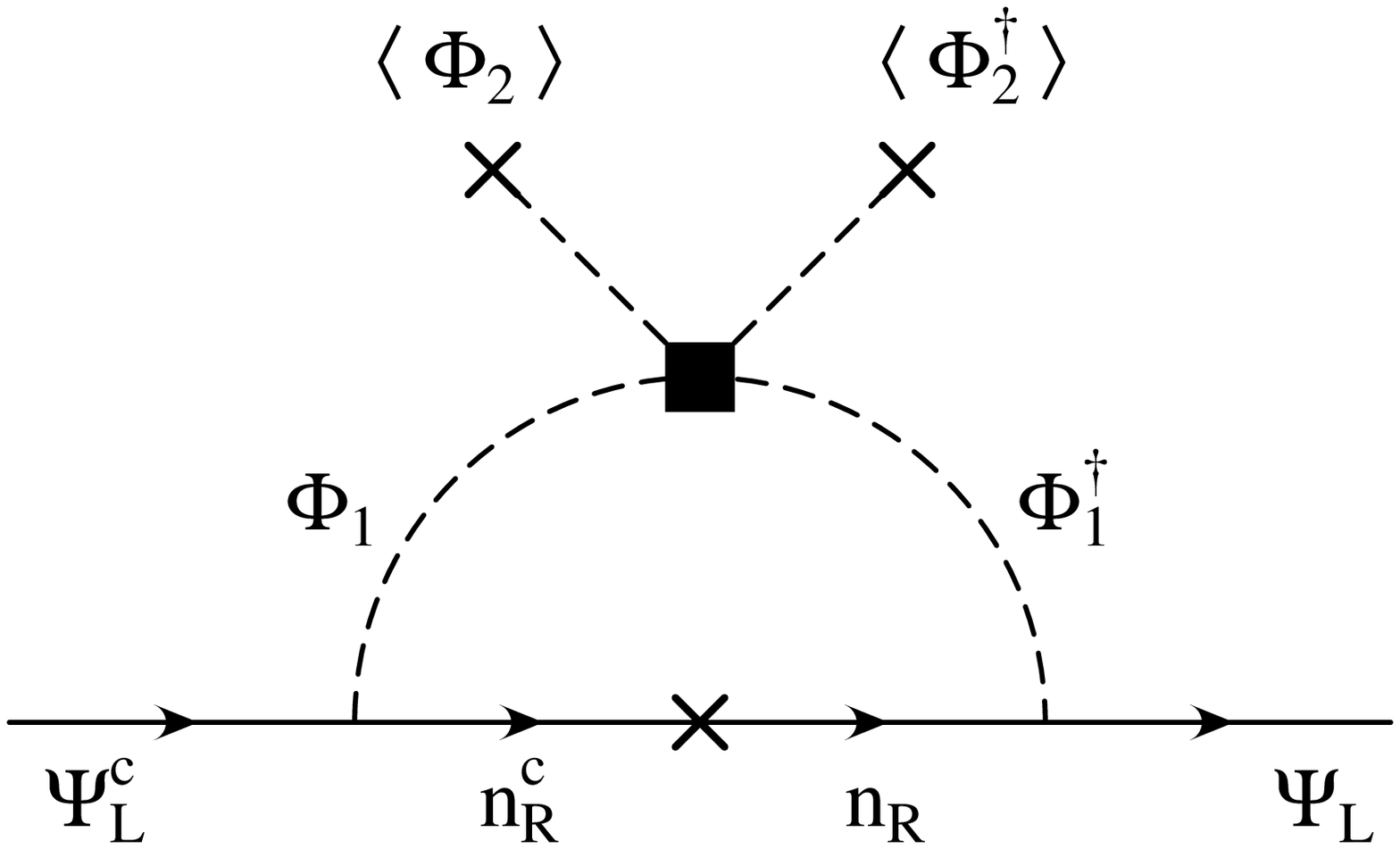,width=2.8in}
\caption{Diagram for Majorana mass matrix of $\Psi_L$.}
\label{fig:majo1}}
First of all, we add a Majorana mass term for $SU(3)_{weak}$ singlet, $n_R$,
to the Lagrangian:
\begin{equation}\label{eq:right_majorana}
-\mathcal L_{\rm Majorana}=m_\diamond\overline{n_R} n^c_R+h.c.,
\end{equation}
whose origin is due to a singlet Higgs or a bare mass term but
is unknown in the context of the simplest little Higgs model.
A UV completion of the model should give an explanation for its presence.
We assert a mass hierarchy, $m_\diamond \ll m_n$, to maintain
the vector-like property of heavy neutral leptons
in the simplest little Higgs model.
Thus the smallness of $m_\diamond$ is expected to be achieved
by quamtum effects in a UV completion of the model.

Next, we introduce Majorana mass term to $SU(2)_L$ singlet $n_L$:
\begin{equation}\label{eq:NMmass}
-\mathcal L_{\rm Majorana}=m_\Box \overline{n^c_L} n_L+h.c.
\end{equation}
Since both $\nu_{eL}$ and $n_L$ belongs to
a $SU(3)_{weak}$ triplet, the presence of 
a Majorana mass term for $n_L$ is expected
to be accompanied with a Majorana mass term for $\nu_{eL}$:
\begin{equation}\label{eq:nuMmass}
-\mathcal L_{\rm Majorana}=m_\nu \overline{\nu^c_{eL}} \nu_{eL}+h.c.
\end{equation}
\FIGURE[l]{\epsfig{file=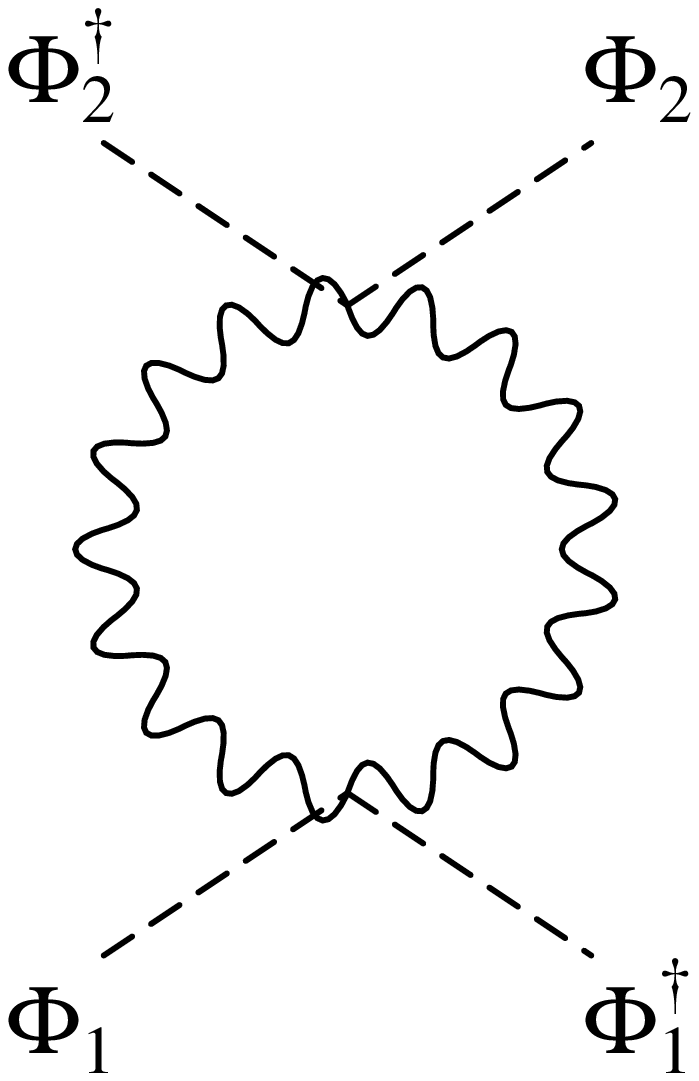,height=1.8in}
\caption{$|\Phi_1^\dagger\Phi_2|^2$ potential arises from gauge loop contributions
which give log-divergent contribution to the Higgs mass.}\label{fig:gauge_loop}}
These Majorana mass terms are achieved at loop level by diagrams
as shown in Fig.~\ref{fig:majo1} where the vertex-cross at bottom
stands for Majorana mass for $n_R$ while the two vertex-crosses at top
represent the vev of $\Phi_2$.
The black square in Fig.~\ref{fig:majo1} represents a coupling
involved with a dimension four operator, 
$|\Phi_1^\dagger\Phi_2|^2$ which gives rise to the
log-divergent contributions to the Higgs mass at one-loop~\cite{Kaplan:2003uc}.
Diagrams in Fig.~\ref{fig:gauge_loop} and Fig.~\ref{fig:fer_loop} show how the operator
$|\Phi_1^\dagger\Phi_2|^2$ is generated at one-loop.
The diagram in Fig.~\ref{fig:gauge_loop} generates a operator of the form:
\begin{equation}
\frac{g^4}{16\pi^2}|\Phi^\dagger_1\Phi_2|^2\log(\Lambda^2/f^2),
\end{equation}
where $\Lambda\sim4\pi f$ is a UV cutoff.
This diagram contributes to the SM Higgs mass term,
$-f^2/16\pi^2 h^\dagger h$,
leading to a lower value of the SM Higgs mass. Therefore, ``$\mu$"-term
is included to enhance the SM Higgs mass parameter:
\begin{equation}
\mathcal L=\mu^2 \Phi^\dagger_1\Phi_2+h.c.
\end{equation}
The diagram in Fig.~\ref{fig:fer_loop} also contributes
to the SM Higgs mass, giving rise to a operator of the form:
\begin{equation}
\lambda_1^2\lambda_2^2/16\pi^2|\Phi_1^\dagger\Phi_2|^2
\log(\Lambda^2/f^2).
\end{equation}
Note that fermion loop contributions are proportional to the two Yukawa
couplings squared and the top quark sector has the largest Yukawa couplongs.
Thus the leading fermion loop contribution comes from the top sector
and all the other quark contributions are safely negligible
compared with the top contribution~\footnote{For the first two generations,
$\lambda_1\ll\lambda_2\sim 1$ in the first model~\cite{Schmaltz:2004de}.}.
In what follows, we sum up all the four $\Phi$'s interactions and 
rewrite it as the effective operator: 
\begin{equation}
\mathcal L=\kappa|\Phi_1^\dagger \Phi_2|^2,
\end{equation}
where $\kappa$ is a dimensionless coupling whose value is
of order $10^{-2}$.

In summary the Majorana masses for the SM neutrino and its heavy partner
are generated at two-loop level and the corresponding low-energy effective
dimension five operator is given by
\begin{equation}\label{eq:Majmass}
\mathcal L_5=
\frac{(\lambda^n)^2}{\Lambda_\nu}
(\Phi_2\overline{\Psi^c_L})(\Phi^\dagger_2\Psi_L)+h.c.,
\end{equation}
with
\begin{equation}
\frac{1}{\Lambda_\nu}
\approx\frac{\kappa}{16\pi^2f^2}m_\diamond.
\end{equation}
After substituting the vev of $\Phi_2$ in Eq.~(\ref{eq:Majmass}),
$\mathcal L_5$ becomes nothing but the Majorana mass matrix of the SM neutrino
and its heavy partner:
\begin{equation}
-\mathcal L_{\rm Majorana}=
\left(\begin{array}{cc} \overline{\nu^c_{eL}} & \overline{n^c_L}\end{array}
\right)
\left(\begin{array}{cc}
\frac{(\lambda^n)^2}{\Lambda_\nu}\frac{v^2}{2}\cos^2\beta &
-\frac{(\lambda^n)^2}{\Lambda_\nu}\frac{vf}{2\sqrt{2}}\sin2\beta \\
-\frac{(\lambda^n)^2}{\Lambda_\nu}\frac{vf}{2\sqrt{2}}\sin2\beta &
\frac{(\lambda^n)^2}{\Lambda_\nu}f^2\sin^2\beta  \end{array}
\right)
\left(\begin{array}{c} \nu_{eL} \\ n_L\end{array}\right)+h.c.
\end{equation}
From Eqs.~(\ref{eq:NMmass}) and~(\ref{eq:nuMmass}) we read that
\begin{equation}\label{eq:Majorana1}
m_\Box \approx \frac{(\lambda^n)^2}{\Lambda_\nu}f^2\sin^2\beta,\qquad
m_\nu \approx \frac{(\lambda^n)^2}{\Lambda_\nu}\frac{v^2}{2}\cos^2\beta.
\end{equation}
Furthermore, there is a mixing term between $\nu_{eL}$ and $n_L$,
whose dimensionful coupling is
\begin{equation}\label{eq:Majorana2}
\hat m\approx-\frac{(\lambda^n)^2}{\Lambda_\nu}\frac{vf}{2\sqrt{2}}\sin2\beta.
\end{equation}
Note that the relative minus sign in Eq.~(\ref{eq:Majorana2})
compared with Eq.~(\ref{eq:Majorana1})
originates from the relative phase difference between $\Phi_1$ and $\Phi_2$,
which is manifest in Eq.~(\ref{eq:two phis})~\footnote{
It will turn on the first-class oscillations in the framework
of the three lepton generation.}.  
\FIGURE[l]{\epsfig{file=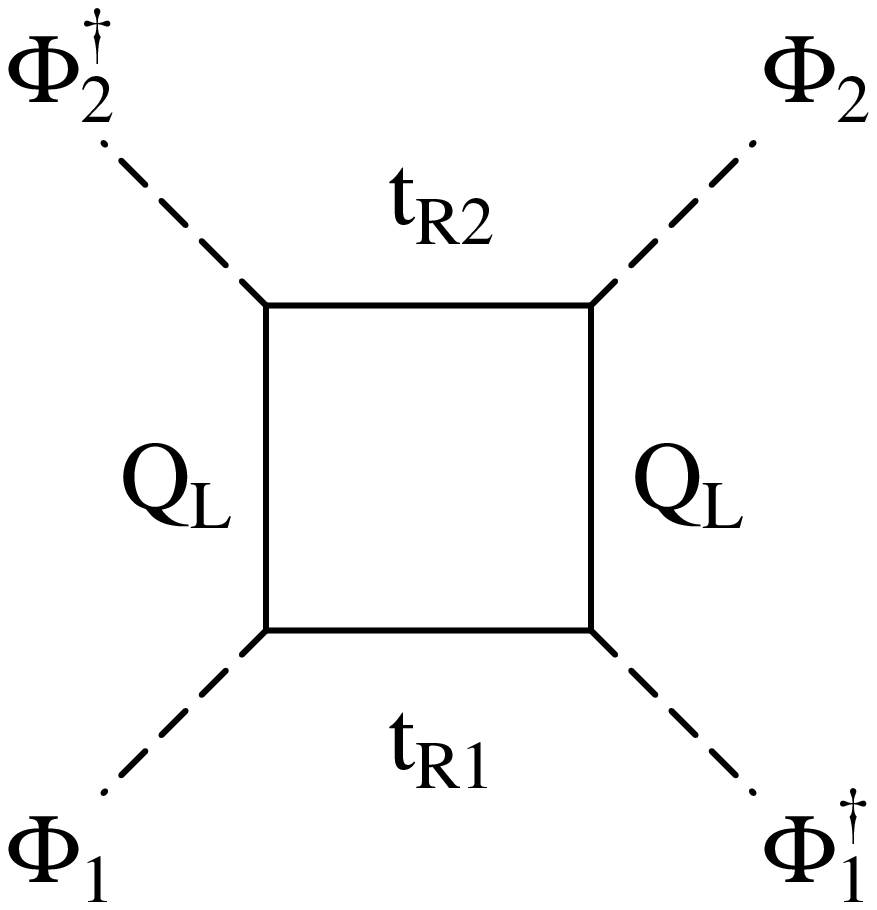,width=1.8in}
\caption{$|\Phi_1^\dagger\Phi_2|^2$ potential arises from fermion loop contributions
which give log-divergent contribution to the Higgs mass.}\label{fig:fer_loop}}

In view of these neutral lepton mass terms, the usual manner of 
proceeding here is to arrange mass matrix for the neutral leptons
in the following $3\times3$ matrix,
\begin{equation}
-\mathcal L_{\rm DM}=
\left(\overline{\nu_L^c}\,,\,\overline{n^c_L}\,,\,\overline{n_R}\right)
{\mathcal M}_{\rm DM}
\left(\begin{array}{c}\nu_L\\ n_L\\ n^c_R\end{array}\right)+h.c.
\end{equation}
with
\begin{equation}
\mathcal M_{\rm DM} \,\equiv\,
\left(\begin{array}{ccc}
\frac{(\lambda^n)^2}{\Lambda_\nu}\frac{v^2}{2}\cos^2\beta &
-\frac{(\lambda^n)^2}{\Lambda_\nu}\frac{vf}{2\sqrt{2}}\sin2\beta &
\frac{\lambda^nv}{\sqrt{2}}\cos\beta \\
-\frac{(\lambda^n)^2}{\Lambda_\nu}\frac{vf}{2\sqrt{2}}\sin2\beta &
\frac{(\lambda^n)^2}{\Lambda_\nu}f^2\sin^2\beta & 
\lambda^nf\sin\beta \\ \frac{\lambda^nv}{\sqrt{2}}\cos\beta &
\lambda^nf\sin\beta & m_\diamond
\end{array}\right).
\end{equation}
Since there is a mass hierarchy among the elements of $\mathcal M$, 
$\lambda^nf>\lambda^n v \gg m_\diamond \gg \{m_\nu,\hat m, m_\Box\}$
we can evaluate physical neutrino mass by computing the product of the
three non-zero eigenvalues of $\mathcal M_{\rm DM}$
\begin{equation}
\mbox{det}\mathcal M_{\rm DM}=-\frac{(\lambda^n)^4f^2 v^2 \sin^22\beta}{2\Lambda_\nu}.
\end{equation}
We compare this result with the product of the two heavy lepton masses
in Eq.~(\ref{new_heavy_ma})
\begin{equation}
\mbox{det}^\prime\mathcal M_{\rm Dirac}=
-(\lambda^n)^2(\frac{v^2}{2}\cos^2\beta+f^2\sin^2\beta).
\end{equation}
To first order in the neutrino mass, the neutral heavy lepton masses are
unchanged, and we end up with
\begin{equation}
m_{\nu_e}=\frac{\mbox{det} \mathcal M_{\rm DM}}{\mbox{det}^\prime \mathcal M_{\rm Dirac}}
=\frac{(\lambda^n)^2}{2\Lambda_\nu}\frac{v^2\sin^22\beta}
{\frac{v^2}{2f^2}\cos^2\beta+\sin^2\beta}.
\end{equation}
From the fact that the heaviest SM neutrino mass is expected to be
of order 0.1 eV,
we can set a rough upper limit on the Majorana mass of $n_R$:
\begin{equation}
m_\diamond\lesssim 1 \times \left[\frac{10^{-2}}{\lambda^n}\right]^2
\left[\frac{10^{-2}}{\kappa}\right]
\left[\frac{f}{1\,\mbox{TeV}}\right]^2 \mbox{GeV},
\end{equation}
which guarantees the previous assumption, $m_\diamond\ll m_n$. It also 
implies that the heavier the mass $m_n$ is the smaller the mass $m_\diamond$ is.

\section{Three generation case}
\label{sec:3g}

Enlarging one generation to three we construct
the lepton Yukawa couplings in three generations.
The general Yukawa couplings for the three generations are given
by extending the coupling numbers in Eq.~(\ref{eq:ecpsi})
to the $3\times3$ coupling matrices:
\begin{equation}\label{eq:ecpsi3}
-\mathcal L_{\rm yuk}=
[\lambda^n]_{ab}\overline{n_{aR}}\Phi^\dagger_1\Psi_{bL}
+\frac{[\lambda^\ell]_{ab}}{f}\overline{\ell_{aR}}
\Phi^i_1\Phi^j_2 \Psi_{bL}^k \epsilon_{ijk} 
+h.c.
\end{equation}
where $\Psi_{aL}=(i\nu_a,i\ell_a,n_a)_L^T$, $n_{aR}$
and $\ell_{aR}$ are a left-handed lepton triplet,
a right-handed neutral singlet and a right-handed charged singlet
under $SU(3)_{weak}$ in the $a$-th generation, respectively.
In what follows square bracket represents a $3\times3$ matix.

We first consider mass eigenstates of the charged leptons by
diagonalizing the $3\times3$ Yukawa coupling matrix, $[\lambda^\ell]$,
which is transformed by two unitary matrices $\mathbf{U}_R$
and $\mathbf{U}_L$:
\begin{equation}
\mathbf{U}_R^\dagger [\lambda^\ell] \mathbf{U}_L =
\mbox{diag}[\lambda^e,\lambda^\mu,\lambda^\tau].
\end{equation}
The charged lepton fields in weak eigenstate, $\ell_L\equiv(\ell_1,\ell_2,\ell_3)_L^T$
and $\ell_R\equiv(\ell_1,\ell_2,\ell_3)_R$, are related to the fields in mass eigenstates,
$\hat \ell_L$ and $\hat \ell_R$ as follows:
\begin{equation}
\ell_R= \mathbf{U}_R \hat\ell_R,\qquad
\ell_L = \mathbf{U}_L \hat\ell_L.
\end{equation}
Thus we find that the three charged leptons have the Dirac masses
as in the SM,
\begin{equation}
(m_e,m_\mu,m_\tau)=\frac{v}{\sqrt{2}}(\lambda^e,\lambda^\mu,\lambda^\tau).
\end{equation}

As for the neutral leptons we consider the first term of the Lagrangian
in Eq.~(\ref{eq:ecpsi3}). We expand it in powers of $1/f$, 
\begin{equation}\label{eq:mass9}
-\mathcal L=
[\lambda^n]_{ab} \overline{n_{aR}} \Phi^\dagger_1 \Psi_{bL}
+h.c. \to
[\lambda^n]_{ab}f\overline{n_{aR}}\left(\sin\beta n_{bL}
+\frac{v}{\sqrt{2}f}\cos\beta \nu_{bL}\right)+h.c.
\end{equation}
The Dirac mass terms for the neutral leptons are present
so that Eq.~(\ref{eq:mass9}) may be rewritten in more suggestive form:
\begin{equation}
-{\mathcal L}_{\rm Dirac}=
\left(\begin{array}{ccc} \overline{\nu^c_L} & \overline{\mathbf{n}^c_L} &
\overline{\mathbf{n}_R}\end{array}\right)
\mathbb{M}_D
\left(\begin{array}{c} \nu_L \\ \mathbf{n}_L \\ \mathbf{n}^c_R \end{array}\right)
+h.c.,
\end{equation}
where the three neutral leptons are 
\begin{equation}
\nu_L=\left(\begin{array}{c}\nu_{1L} \\ \nu_{2L} \\ \nu_{3L}\end{array}\right),\qquad
\mathbf{n}_L=\left(\begin{array}{c}n_{1L} \\ n_{2L} \\ n_{3L}\end{array}\right),\qquad
\mathbf{n}_R=\left(\begin{array}{c}n_{1R} \\ n_{2R} \\ n_{3R}\end{array}\right),
\end{equation}
and the $9\times9$ Dirac mass matrix is
\begin{equation}
\mathbb{M}_D\equiv\left(\begin{array}{ccc} 
[0] & [0] & [\lambda^n]^\dagger\frac{v}{\sqrt{2}}\cos\beta \\
{[0]} & [0] & [\lambda^n]^\dagger f\sin\beta \\
{[\lambda^n]} \frac{v}{\sqrt{2}}\cos\beta
& [\lambda^n] f \sin\beta & [0] \end{array}\right).
\end{equation}
We diagonalize the mass mastrix $\mathbb{M}_D$
by a unitary transformation:
\begin{equation}
\mathbb{W}_0^\dagger \mathbb{M}_D\mathbb{W}_0=\mathbb{M}^{\rm diag},
\end{equation}
where a transformation matrix $\mathbb{W}_0$ is given by
\begin{equation}\label{eq:transf matrix}
\mathbb{W}_0=\left(\begin{array}{ccc}
\cos\theta[1] & \sin\theta[1] & [0]\\
-\sin\theta[1] & \cos\theta[1] & [0] \\
{[0]} & [0] & [1]\end{array}
\right)
\left(\begin{array}{ccc} [1] & [0] & [0] \\
{[0]} & \mathbf{V}_L & [0] \\
{[0]} & [0] & \mathbf{V}_R\end{array}\right) 
\left(\begin{array}{ccc} [1] & [0] & [0] \\
{[0]}& -\frac{1}{\sqrt{2}}[1] & \frac{1}{\sqrt{2}}[1] \\
{[0]} & \frac{1}{\sqrt{2}}[1] & \frac{1}{\sqrt{2}}[1]
\end{array}\right),
\end{equation}
 with $\mathbf{V}_R$ and  $\mathbf{V}_L$ being $3\times3$ unitary transformation matrices
 which diagonalize $[\lambda^n]$: 
\begin{equation}
\mathbf{V}_R^\dagger [\lambda^n]\mathbf{V}_L=\mbox{diag}[\lambda^n_1,
\lambda^n_2,\lambda^n_3].
\end{equation}
The diagonal matrix $\mathbb{M}^{\rm diag}$ is given by
\begin{equation}
\mathbb{M}^{\rm diag}=\mbox{diag}[0,0,0,-\lambda^n_1,-\lambda^n_2,-\lambda^n_3,
\lambda^n_1,\lambda^n_2,\lambda^n_3].
\end{equation}
Here we, for simplicity, assume that $\lambda^n_1$, $\lambda^n_2$
and $\lambda^n_3$ are all real and further are all different, 
$\lambda^n_1\neq\lambda^n_2\neq\lambda^n_3$. Note that the first three zeros
in $\mathbb{M}^{\rm diag}$ imply massless SM neutrinos while the remaining
three pairs of mass eigenvalues give rise to Dirac masses for the three
heavy neutral leptons. Using Eq.~(\ref{eq:transf matrix}) we can write the mass
eigenstates $(\hat \nu_L,\mathbf{\hat n}_L, \mathbf{\hat n}_R^c)$ in the linear
combinations of the weak eigenstates $(\nu_L,\mathbf{n}_L, \mathbf{n}_R^c)$:
\begin{equation}\label{eq:linear comb}
\left(\begin{array}{c}
\hat \nu_L \\ \mathbf{\hat n}_L \\  \mathbf{\hat n}_R^c\end{array}\right)
=\left(\begin{array}{ccc} \cos\theta[1] & -\sin\theta[1] & [0] \\
-\frac{1}{\sqrt{2}} \mathbf{V}_L\sin\theta[1] &-\frac{1}{\sqrt{2}} \mathbf{V}_L\cos\theta[1]
& \frac{1}{\sqrt{2}}\mathbf{V}_R[1] \\
\frac{1}{\sqrt{2}} \mathbf{V}_L\sin\theta[1] &\frac{1}{\sqrt{2}} \mathbf{V}_L\cos\theta[1]
& \frac{1}{\sqrt{2}}\mathbf{V}_R[1] \end{array}\right)
\left(\begin{array}{c}
\nu_L \\ \mathbf{n}_L \\  \mathbf{n}_R^c\end{array}\right).
\end{equation}

In order to make the SM neutrino acquire masses we include the Majarana mass terms
for the neutral leptons to the Lagrangian in Eq.~(\ref{eq:ecpsi3})
as in Section \ref{sec:one}.
As we enlarge one generation to three the Majorana masses 
in Eqs.~(\ref{eq:right_majorana}), (\ref{eq:Majorana1}) and (\ref{eq:Majorana2})
are replaced with $3\times3$ Majorana mass matrices, respectively:
\begin{equation}
-\mathcal L_{\rm Majorana}=
\left(\begin{array}{ccc}\overline{\nu^c_L} & \overline{\mathbf{n}^c_L}
& \overline{\mathbf{n}_R}\end{array}\right)
\mathbb M_M
\left(\begin{array}{c}\nu_L \\ \mathbf{n}_L \\ \mathbf{n}^c_R\end{array}\right)
+h.c.,
\end{equation}
where the $9\times9$ Majorana mass matrix is
\begin{equation}\label{eq:Majorana 99}
\mathbb M_M=
\left(\begin{array}{ccc}
[m_\nu] & [\hat m]^\dagger & [0]\\
{[\hat m]} & [m_\Box] & [0]\\
{[0]} & [0] & [m_\diamond]\end{array}\right).
\end{equation}
Here we make a simplifying assumption that $[m_\diamond]=m_\diamond\mbox{diag}[1,1,1]$.
This allows us to present our results in more suggestive form.
But it does not affect in any essential way the physics which we shall draw.
With the assumption we can write the $3\times3$ Majorana mass matrices
in Eq.~(\ref{eq:Majorana 99}) as follows:
\begin{eqnarray}
[m_\nu]&\approx&
\frac{[\lambda^n]^\dagger[\lambda^n]}{\Lambda_\nu}\frac{v^2}{2}\cos^2\beta,
\label{eqs:33masses1}\\
{[m_\Box]}&\approx& \frac{[\lambda^n]^\dagger[\lambda^n]}{\Lambda_\nu}f^2\sin^2\beta,
\label{eqs:33masses2} \\
{[\hat m]}=[\hat m]^\dagger&\approx&
-\frac{[\lambda^n]^\dagger[\lambda^n]}{\Lambda_\nu}\frac{vf}{2\sqrt{2}}\sin 2\beta.
\label{eqs:33masses3}
\end{eqnarray}
Note that all the three matrices are proportional to $[\lambda^n]^\dagger[\lambda^n]$
but differ only in vev's. Now the full $9\times9$ mass matrix is given by
$\mathbb{M}_D+\mathbb{M}_M$.
Comparing with $\mathbb{M}_D$, we observe that the elements in $\mathbb{M}_M$
are much smaller than those in $\mathbb{M}_D$, so that the perturbation theory
can be applied to approximately compute the nonzero SM neutrino masses.
Accordingly $\mathbb M_D$ and $\mathbb M_M$ act as the {\it unperturbed} and
{\it perturbed} mass matrix in the perturbation theory, respectively.
However, since there still remain threefold degenerate zero eigenvalues,
we in particular use the perturbation method for the degenerate case.

Let us carry out the perturbation for the degenerate case.
First of all, we need to find the projection operator  $P_0$ onto the three SM neutrino
states among the nine neutral lepton states.
From Eq.~(\ref{eq:transf matrix}), $P_0$ is given by
\begin{equation}
P_0=\left(\begin{array}{ccc}
\cos^2\theta[1] & -\cos\theta\sin\theta[1] & [0] \\ -\cos\theta\sin\theta[1] & \sin^2\theta[1] & [0] \\
{[0]} & [0] & [0]\end{array}\right).
\end{equation}
As a result, the mass eigenvalues of the SM neutrinos to the first order
are just the nonzero roots of the characteristic equation in variable
$\Delta$,
\begin{equation}
\mbox{det}[P_0 \mathbb M_M P_0-\Delta\mathbb I\,]=0,
\end{equation}
where $\mathbb I\,$ is the identity matrix.
From Eqs.~(\ref{eqs:33masses1}), (\ref{eqs:33masses2}) and (\ref{eqs:33masses3}),
the nonzero roots of the characteristic equation are
\begin{equation}\label{eq:3nm}
\Delta_i=m^\nu_i=
\frac{(\lambda^n_i)^2}{2\Lambda_\nu}\frac{v^2\sin^22\beta}{\sin^2\beta+
\frac{v^2}{2f^2}\cos^2\beta}\qquad
\mbox{with}\quad i=1,2,3, 
\end{equation}
which are the SM neutrino masses to the {\it first} order.
Remind that the neutrino are massless to the {\it zeroth} order in the perturbation theory.
In addition, the characteristic equation gives the mixing angles for the SM neutrinos
to the {\it zeroth} order 
\begin{equation}\label{eq:flc}
\hat \nu_L=\mathbf{V}_L(\cos\theta \nu_L-\sin\theta \mathbf{n}_L).
\end{equation}
From Eq.~(\ref{eq:3nm}), it is straightforward to see that the SM neutrino masses
are proportional to $(\lambda_i^n)^2$, yielding a mass relation between the SM neutrinos
and heavy neutral leptons as follows:
\begin{equation}
m^\nu_1:m^\nu_2:m^\nu_3\approx m^2_{n_1}:m^2_{n_2}:m^2_{n_3}.
\end{equation}
This implies that the neutrino mass hierarchy is revealed by
the heavy neutral lepton mass hierarchy. Therefore, measuring of the heavy neutral
lepton masses at the International Linear Collider (ILC) can tell the neutrino mass
hierarchy that is to be disclosed at the neutrino experiments in the near future.

Using Eq.~(\ref{eq:linear comb}) and (\ref{eq:flc}), we can construct 
a full $9\times9$ unitary leptonic mixing matrix to the {\it zeroth} order
\begin{equation}\label{eq:99mza}
\mathcal W_0=
\left(\begin{array}{ccc} 
\mathbf{V}^\dagger_L\cos\theta[1] & -\frac{1}{\sqrt{2}} \mathbf{V}^\dagger_L\sin\theta[1] &
\frac{1}{\sqrt{2}} \mathbf{V}^\dagger_L\sin\theta[1] \\
-\mathbf{V}^\dagger_L\sin\theta[1] & -\frac{1}{\sqrt{2}} \mathbf{V}^\dagger_L\cos\theta[1] &
\frac{1}{\sqrt{2}} \mathbf{V}^\dagger_L\cos\theta[1] \\
{[0]} & \frac{1}{\sqrt{2}}\mathbf{V}^\dagger_R[1]&\frac{1}{\sqrt{2}}\mathbf{V}^\dagger_R[1] \end{array}\right).
\end{equation}
Thus the weak eigenstates of the neutral leptons are given
by the linear combinations of the mass eigenstates
\begin{equation}
\left(\begin{array}{c} \nu_L \\ \mathbf{n}_L \\ \mathbf{n}^c_R \end{array}\right)= \mathcal W_0 
\left(\begin{array}{c} \hat \nu_L \\ \mathbf{\hat n}_L \\ \mathbf{\hat n}^c_R \end{array}\right).
\end{equation}
From Eq.~(\ref{eq:99mza}), the MNS mixing matrix is given as
\begin{equation}
\mathcal U_{MNS} =\cos\theta\mathbf{V}^\dagger_L,
\end{equation} 
which is obviously {\it not} unitary due to $\cos\theta\neq1$.
Furthermore, neutrino-heavy neutral
lepton mixing angles are $\sin\theta\mathbf{V}^\dagger_L$ which is obviously not small.
Does it conflict with the experimental data on the neutrino oscillations? It is not.
In fact, we can not extract the value of $\cos\theta$ from the neutrino oscillation experiments.
Due to the tremendous mass difference between the SM neutrinos and the heavy neutral
leptons, such a mixing yields heavy neutral lepton decays 
rather than neutrino-heavy neutral lepton oscillations.

One can further take into account leptonic weak charged currents
so as to extract the value of $\cos\theta$.
But the mixing angles in the quark sector exhibit the same patterns because  
the Yukawa couplings in the quark sector are of the same form as those in the lepton sector.
Thus we can not see the presence of $\cos\theta$ in leptonic weak decays associated with the SM neutrinos by comparing leptonic charged weak currents with quark charged weak currents, i.e.
in the weak coupling ratio, $(ud\to e\nu_e)$ /$(\mu\nu_\mu\to e\nu_e)$, $\cos\theta$
cancels out.

Finally, we can extract the angle $\cos\theta$ (or $\sin\theta$) from study
of the heavy neutral lepton decays. Classifying various decay modes of the heavy neutral leptons
we also find $\mathbf{V}^\dagger_L$. With these results, one can confirm the MNS mixing matrix
that is obtained at the neutrino oscillations.
 
\section{Summary and outlook}

We have analyzed the lepton sector in the simplest little Higgs model which contains
vector-like neutral leptons as heavy partners to the SM neutrinos.
We emphasize that a naive construction of the neutral lepton sector at tree level
yield a large Dirac mass to the heavy neutral leptons but no masses to the SM neutrinos.
We have shown that one loop diagrams involved with two Higgs triplets yield 
small Majorana masses for the neutral leptons including the SM neutrinos,
leading to not only neutrino flavor mixing but also neutrino-heavy neutral lepton mixing.

We have evaluated a $9\times9$ mixing matrix of the neutral leptons and
the SM neutrino masses by the standards of perturbation theory. The ratio of the
Majorana masses associated with the neutral leptons to the Dirac masses is so small
that the higher order terms of the perturbation theory can be negligible.
We therefore feel that our computations, though a rough approximation, does not change
the outcomes that we have drawn. 
We can acknowledge that the heavy neutral leptons are directly associated with the SM neutrinos
in the simplest little Higgs model and thus the study of the heavy neutral leptons catches
a glimpse of the SM neutrinos and vice versa.
For example, the mass hierarchy of the heavy neutral leptons, once revealed at the ILC,
will back up that of SM neutrinos which will be found at the neutrino oscillation experiments
in the near future. Inversely, using the currently known data on the neutrinos one can predict or
restrict the properties of heavy neutral leptons.

There are many paths for future research on the neutrinos in the simplest little Higgs model.
One can further investigate phenomena associated with the neutral lepton mixing, such as the various decay modes of the heavy neutral leptons. One can boldly explore the origin of Majorana mass
for the $SU(3)_{weak}$ singlet in a UV completion of the simplest little Higgs model. 
Finally, with the help of our scenario one can investigate the neutrinos in other little Higgs
models which embrace heavy neutral leptons, such as different simple little Higgs model based on a larger simple group.
\bigskip

\acknowledgments

I am grateful to Ann Nelson for instructive comments, in particular, recommending me to use the perturbation theory for computation of the mixing matrix.
I would like to express thanks to Sukjin Yoon for answering to my questions on matrix computation.

%


\end{document}